\documentclass[10pt,letterpaper]{article}
\usepackage{amsmath}
\usepackage{graphicx}
\usepackage{textgreek}
\begin{document}
\graphicspath{{C:/Users/nicolas/Dropbox/FlowLimitations/Paper/figs/eps/}}
\title{Simultaneous measurement of localized diffusion and flow using optical coherence tomography}

\author{Nicol\'as Weiss$^{1}$, Ton G. van Leeuwen$^{1}$, and Jeroen Kalkman$^{2}$}

\newcommand{\Addresses}{{
  \medskip
  \footnotesize
  \textsc{$^{1}$Biomedical Engineering \& Physics, Academic Medical Center, University of Amsterdam, PO Box 22700, 1100 DE Amsterdam, the Netherlands}\par\nopagebreak
 
  \medskip

  \textsc{$^{2}$Department of Imaging Physics, Faculty of Applied Sciences, Delft University of Technology, Lorentzweg 1, 2628 CJ Delft, the Netherlands}\par\nopagebreak
\center \texttt{n.m.weiss@amc.uva.nl}

}}

\maketitle
\Addresses
\begin{abstract}We report on the simultaneous and localized measurements of the diffusion coefficient and flow velocity based on the normalized autocorrelation function using optical coherence tomography (OCT). Our results on a flowing solution of polystyrene spheres show that the flow velocity and the diffusion coefficient can be reliably estimated in a regime determined by the sample diffusivity, the local flow velocity, and the Gaussian beam waist. We experimentally show that a smaller beam waist results in an improvement of the velocity sensitivity at cost of the precision and accuracy of the estimation of the diffusion coefficient. Further, we show that the decay of the OCT autocorrelation due to flow depends only on the Gaussian beam waist irrespective of the sample position with respect to the focus position.   
\end{abstract}


\section{Introduction}
Quantification of diffusive and translational dynamics of particles is interesting for the study of fundamental fluid dynamic processes such as, shear dependent diffusion \cite{Orihara2011}, and for a number of applications such as e.g., aerosols \cite{Weber1993}, particle sorting \cite{DiCarlo2007}, intracellular transport \cite{Joo2010}, biofilm growth \cite{Drescher2013}, and blood circulation \cite{Lee2013}. 
%
%

In principle two techniques are available to quantify flow and diffusion simultaneously. First, particle tracking based techniques such as particle tracking velocimetry (PTV), use a sequence of (microscopy) images to track the motion of individual particles which are then used to calculate the particle displacement to find the relevant dynamic parameters. Second, dynamic light scattering (DLS) based technique detect the fluctuations of single scattered light by an ensemble of particles without resolving the individual particles. By fitting an appropriate model to the autocorrelation function of the scattered light, the ensemble averaged particle dynamics are measured. However, in the case of PTV the focus has to be mechanically scanned along the propagation direction of the light to produce useful imaging ranges and in DLS the path-length distribution of the scattered light is not known, providing only volumetric averaged information about the ensemble particle dynamics.

Localized particle dynamics can be probed using optical coherence tomography (OCT). Using a combination of coherent and confocal gating OCT measures the exact path-length distribution of the scattered light up to a few millimeters deep into a sample and with high spatio-temporal resolution \cite{Huang1991}. Functional extensions based on the combination of OCT and DLS have been developed to quantify particle diffusion \cite{Bizheva1998, Kalkman2010, Ishii2010, Chhetri2011}, flow velocity \cite{Huang2014, Uribe2014}, and flow and diffusion \cite{Lee2012, Weiss2013}.     

In this manuscript we report on simultaneous and localized measurements of the diffusion coefficient and flow velocity of a colloidal suspension. We quantify the fluctuations in the measured backscattered intensity by fitting a model to the normalized autocorrelation function. Previously, we have shown that the autocorrelation function can be used to quantify the flow velocity for arbitrary flow angles in the presence of diffusion \cite{Weiss2013}. In the case of shear flow the velocity gradient over the coherent detection gate is a source of additional decorrelation of the OCT signal. 
With the aim of studying the coupling of the flow and diffusion process in the measurement model, we constrain the measurement to be perpendicular to the flow direction. Although usually, both processes are assumed to be independent sources of decorrelation, we show that the flow velocity and the diffusion coefficient can only be reliably estimated in a regime determined by the sample diffusivity, the local flow velocity, and the Gaussian beam waist.
\section{Materials and Methods}
\subsection{Optical coherence tomography system}
\begin{figure}[t]
	\centering
		\includegraphics[width=0.6\textwidth]{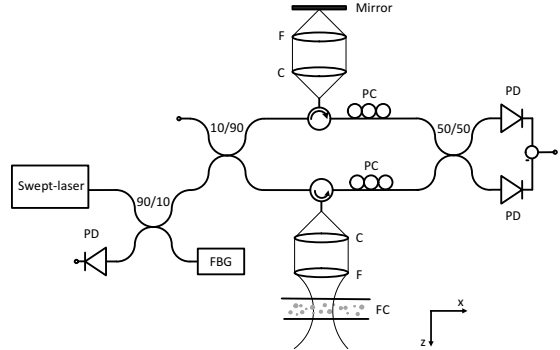}
\caption{Schematic of the experimental swept-source OCT set-up. PD: photodetector, FBG: fiber Bragg grating, PC: polarization controllers, C: collimating lens, F: focusing lens, and FC: flow channel. Gravity is in the $z$-direction. Adapted from Ref. \cite{Weiss2013}.}
\label{fig:setup}
\end{figure}

The experiments are performed with a home built fiber-based swept-source OCT system. A schematic of the experimental set-up is shown in Fig. \ref{fig:setup}. The description of the system has been reported elsewhere \cite{Weiss2013} but is repeated here for convenience. The system operates at a center wavelength of $1312$ nm with a bandwidth of $92$ nm and a sweep frequency of $50$ kHz (Axsun Technologies). The average output power is 20.9 mW and the duty cycle is 59.4\%. Data is sampled (ATS9350, AlazarTech) with an interferometrically derived external clock signal at equidistant wavenumber intervals. To ensure phase stability each sweep is triggered by the signal of a fiber Bragg grating centered at $1266$ nm (OE Land) \cite{Hendargo2011}. The interferometric signal is detected with a 150 MHz balanced photodetector (PDB450C, Thorlabs) and a 80 MHz low-pass filter (VLF-80+, Mini-Circuits). The trigger signal is detected with a 125 MHz photodetector (1811, New Focus). To assess the influence of the focusing optics on the measured parameters, the sample and reference arms' optics are composed of a collimating lens (PAF-X-18-C, Thorlabs) and three different achromatic doublet focusing lenses (AC254-030-C, AC254-040-C, AC254-100-C, Thorlabs). Unless otherwise stated in the text the $40$ mm focal length lens (AC254-040-C) is used. The power ratio of the sample and reference arms is $90/10$. The axial resolution is $8.1 \pm 0.3$ \textmu m in air measured with a mirror reflector. 
%
%
%
%
\subsection{Flow system} 
Flow is generated by a perfusion pump (Perfusor fm, Braun) and directed through a cylindrical glass channel with a measured inner diameter of $1097 \pm 25$ \textmu m or a glass channel with an inner diameter of $50 \pm 5$ \textmu m (VitroCom). The flowing suspension consists of 1 vol.~$\%$ polystyrene spheres (PPs-$0.2$, Kisker) dissolved in distilled water. The measured mean sphere diameter is $152 \pm 15$ nm (qNano, np400, Izon). The measured refractive index of the medium at $1312$ nm is $n = 1.33$. The flow channel is placed perpendicular to the propagation direction of the imaging beam by minimizing the Doppler shift. The experimental flow conditions throughout the manuscript are well described by Poiseuille flow with a maximum Reynolds number of $9$.
\subsection{Data analysis}
The path-length resolved diffusion coefficient and flow velocity are measured by fitting a model of the normalized autocorrelation function of the OCT signal for every path-length independently using a modified version of the model presented in Ref. \cite{Weiss2013}:   
\begin{equation}
g(z,\tau) = e^{-2D(z)q^2|\tau|} e^{- 2\left[ \frac{v(z)\tau}{w} \right]^2}  \ ,
\label{eq:g}
\end{equation}
where the exponential term describes the longitudinal diffusive dynamics and the Gaussian term describes the transverse directional dynamics, with $z$ representing the optical path-length (OPL) and $|\tau|$ the time. The additional factor of $2$ in the exponents accounts for the use of the magnitude of the OCT signal in the analysis \cite{Kalkman2010}. $D(z)$ is the path-length resolved diffusion coefficient given by the Stokes-Einstein equation $D(z) = k_BT_K / 6 \pi \eta r$, with $k_B$ Boltzmann's constant, $T_K$ the absolute temperature, $\eta$ the viscosity, and $r$ the hydrodynamic particle radius. The absolute value of the scattering vector is $q = 4\pi n \sin{(\alpha/2)}/\lambda$, with $n$ the refractive index of the medium, $\lambda$ the wavelength in vacuum, and $\alpha$ the scattering angle. Further, $v(z)$ is the path-length resolved transverse flow velocity and $w$ is the beam waist ((1/e) radius of the field). The normalization was taken with respect of $g(z,0)$. For our OCT set-up, the spread of $q$ over the bandwidth is small, therefore we set $q = q_c$ at the center wavelength and $\alpha = 180^\circ$. For the derivation of Eq. \ref{eq:g} we have assumed no number fluctuations, independence of particle concentration, and single scattering \cite{Weiss2013}.
\subsection{Model fitting}
Processing of the data is performed as follows: raw interferometric data consisting of 1088 data points is Fourier transformed to calculate the complex-valued OCT signal. For every path-length we calculate the autocovariance of the magnitude of the OCT signal over 5000 time-adjacent acquisitions. This process is repeated 20 times and averaged. The transverse velocity and diffusion coefficient are determined in the time domain by fitting Eq.~(\ref{eq:g}) to the normalized autocovariance of the data. In this fit, $v(z)$ and $D(z)$ are the only free running parameters. In all plots throughout the manuscript, all values of $v$ and $D$ are mean values over 5 measurements and the error bars are the corresponding standard deviations. The error on the fitted parameters is expressed as the coefficient of variation, which is defined as the ratio of the standard deviation to the mean. Throughout the manuscript the results of the model fitting are shown in the frequency domain. 
\section{Results}
\subsection{Diffusion}
\label{sec:noflow}
\begin{figure}[t]
	\centering
		\includegraphics[width=0.7\textwidth]{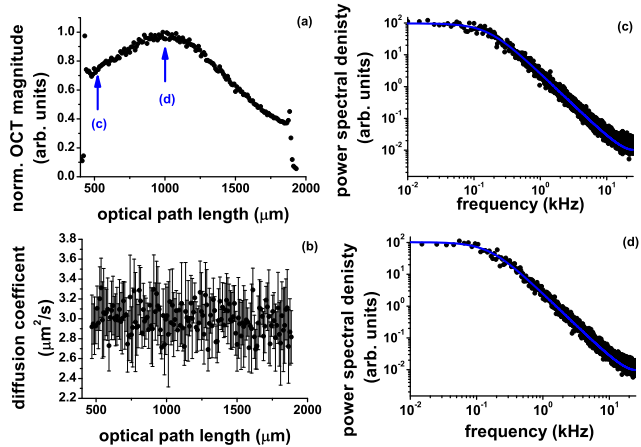}
\caption{(Color online) Measurement of the local diffusion coefficient of the polystyrene solution in the absence of flow: (a) Plot of the normalized OCT magnitude of the flow channel. (b) Path-length resolved diffusion coefficient for the polystyrene solution. (c) and (d) Log-log plot of the power spectral density for the optical path-lengths shown by the arrows in (a). The circles represent the measured data and the blue line represents the model including only the diffusion term.}
\label{fig:static}
\end{figure}

Figure \ref{fig:static} shows results for the measurement of the diffusion coefficient for a solution of polystyrene spheres in the absence of flow. Figure \ref{fig:static}(a) shows a typical plot of the normalized OCT magnitude for the flow channel. The focus was placed near the center of the channel to reduce the influence of multiple scattering at the longer path-lengths \cite{Weiss2013}. Figure \ref{fig:static}(b) shows the path-length resolved diffusion coefficient for the polystyrene solution. The calculated mean diffusion coefficient over all path-lengths is  $3.0 \pm 0.2$ \textmu m$^2$/s. This is in good agreement with the expected diffusion coefficient of $2.8 \pm 0.3$ \textmu m$^2$/s. It should be noted that boundary effects on the diffusion are to be expected, however the range of distances at which these effects occurs is small compared to the coherence length of the laser source \cite{Ishii2010}. The power spectral density at path-lengths of $516$ \textmu m and of $994$ \textmu m is calculated and plotted in Fig. \ref{fig:static}(c) and (d), respectively. As can be observed, the data (circles) and the model (line) are in good agreement. The small deviation form the model at the end of the frequency range is attributed to noise.
\subsection{Flow and diffusion, one parameter fit} \label{sec:fixeddiff}
\begin{figure}[t]
	\centering
		\includegraphics[width=0.8\textwidth]{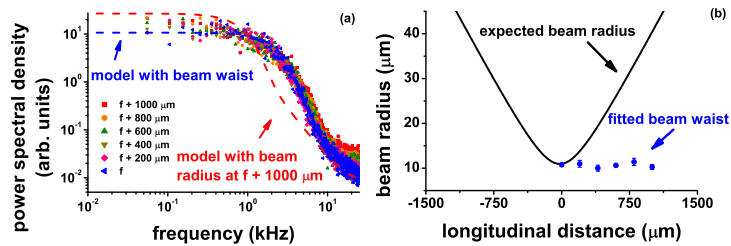}
\caption{(Color online) Independence of the autocorrelation function from the local beam radius: (a) Power spectral densities measured at the center of a $50$ \textmu m in diameter flow channel at different distances from the focus but for the same flow velocity. The markers show the measurements and the dashed lines show the expected models using the beam waist (blue) and using the beam radius at $1000$ \textmu m away from the focus (red). (b) The line shows the expected beam radius for a focused Gaussian beam with a focal length of 40 mm. The markers show the fitted beam waist by letting $w$ be the only fitting parameter in Eq. (\ref{eq:g}).}
\label{fig:RadiusIndependence}
\end{figure}

We start by showing that the decay of the Gaussian term in the normalized OCT autocorrelation function in Eq. (\ref{eq:g}) is characterized by the beam waist and not by the local beam radius irrespective of sample position from the focus. Figure \ref{fig:RadiusIndependence} (a) shows plots of the power spectral density for an experiment where a $50$ \textmu m diameter channel was translated in the longitudinal direction away from the position of the focus. The markers show the data corresponding to different distances from the focus, but with the same flow velocity. The data is compared to a model that depends only on the beam waist at the focus and to a model that is based on a depth dependent beam radius. As can be clearly seen, all power spectra overlap and it can be observed that the decay of the power spectral density does not depend on the value of the beam radius at the corresponding distance from the focus, but depends only on the beam waist. The deviations from the model for high frequencies is attributed to the influence of noise which increases with increasing distance from the focus. Figure \ref{fig:RadiusIndependence} (b) shows the expected beam radius for a focused Gaussian beam for a lens with a focal length of 40 mm and the values for the beam waist calculated by fitting the data shown in (a) and by letting $w$ be the only fitting parameter in Eq. (\ref{eq:g}). Clearly, the fitted beam waist values are independent of the position of the sample with respect to the focus.   

\begin{figure}[t]
	\centering
		\includegraphics[width=0.7\textwidth]{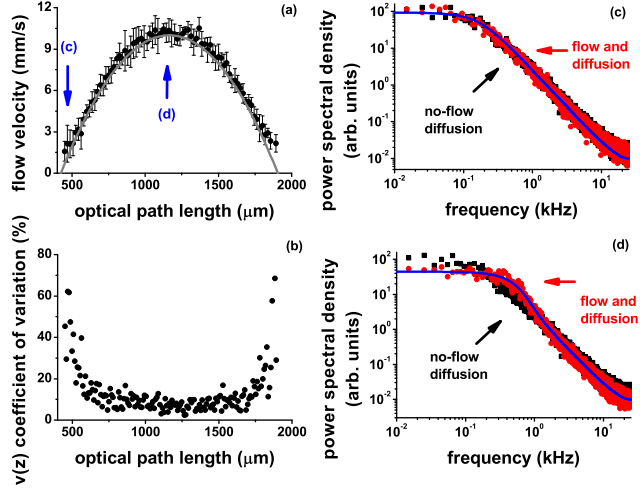}
\caption{(Color online) (a) Path-length resolved flow velocity through the cylindrical flow channel. The gray parabola shows the reference velocity. The arrows correspond to the path-lengths shown in (c) and (d). (b) Coefficient of variation for the flow velocities shown in (a). (c) Power spectral density for a path-length close to the wall of the channel, the red circles show the measurement and the blue line shows the fitted model. For comparison, the black squares show the data for the no-flow diffusion case (cf. Fig. \ref{fig:static}(b)). (d) Similar to (c), but for a path-length corresponding to the center of the flow channel. For the sake of visualization, only every other data point has been plotted.}
\label{fig:TransFlow}
\end{figure}

Figure \ref{fig:TransFlow} shows results for the measurement of the path-length resolved flow velocity with the diffusion coefficient fixed to the value measured in the no-flow case (cf. Sec. \ref{sec:noflow}). Since the diffusion coefficient does not depend on optical path-length, $v(z)$ is the only fitting parameter in Eq. (\ref{eq:g}). Figure \ref{fig:TransFlow}(a) shows the path-length resolved flow velocity. The gray parabola shows the reference velocity calculated based on the flow rate set by the perfusion pump and the diameter of the channel. As can be observed, the computed flow velocity is in good agreement with the reference values. Figure \ref{fig:TransFlow}(b) shows the coefficient of variation of the fitted flow velocity. For optical path-lengths close to the walls of the channel an increase in the coefficient of variation is observed. Figure \ref{fig:TransFlow}(c) and (d) show plots of the power spectral density for the path-lengths shown by the arrows in Fig. \ref{fig:TransFlow}(a). The circles show the data and the squares show the data corresponding to the no-flow diffusion experiment. For the path-length close to the wall of the channel, the decay of the power spectral density for the flow experiment is well described by a Lorentzian decay as was the case for the no-flow experiment. However, for the path-length corresponding to the center of the channel, this is no longer the case: the contribution of the Gaussian (flow) term is clearly observed in the first half of the frequency range. For all path-lengths, a excellent agreement is observed between the data and the model. 

\begin{figure}[t]
	\centering
		\includegraphics[width=0.7\textwidth]{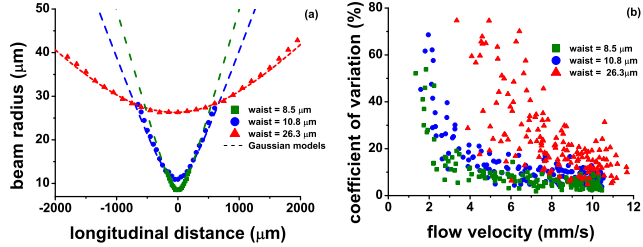}
\caption{(Color online) (a) Beam radius in the propagation direction of the imaging beam for three focusing lenses measured by a knife edge. The markers represent the data and the dashed lines show the corresponding Gaussian beam models. (b) Coefficient of variation for the measured flow velocity.}
\label{fig:RelError}
\end{figure}

Figure \ref{fig:RelError} shows results for the dependency of the velocity uncertainty from Gaussian beam waist. Figure \ref{fig:RelError}(a) shows a plot of the beam radius for three focusing lenses measured with a knife edge \cite{Arnaud1971}. The squares correspond to a lens with a focal length of $30$ mm, the circles for a lens with a focal length of $40$ mm, and the triangles for a lens with a focal length of $100$ mm. The dashed lines show the corresponding Gaussian beam models. The measured waist values are $8.5 \pm 0.5$ \textmu m, $10.8 \pm 0.5$ \textmu m, and $26.3 \pm 1$ \textmu m, respectively. Figure \ref{fig:RelError}(b) shows the flow velocity's coefficient of variation measured in three different experiments with the three lenses. For each experiment the maximum reference flow velocity is $10$ mm/s. As can be observed, the coefficient of variation monotonically decreases for increasing flow velocity for all three lenses. For a particular flow velocity, the choice of a smaller waist results in a lower value for the coefficient of variation compared to the values measured with a larger waist. 


\subsection{Flow and diffusion, two parameter fit}
\begin{figure}[t]
	\centering
		\includegraphics[width=0.7\textwidth]{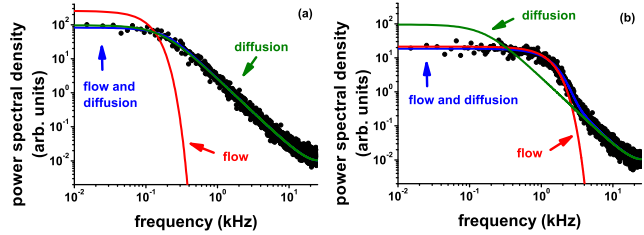}
\caption{(Color online) Power spectral densities for an optical path-length close to the edge of channel (a) and for an optical path-length at the center of the channel (b). The circles show the data, the blue line shows the full model accounting for flow and diffusion, the red line shows the model accounting only for flow, and the green line shows the model accounting only for diffusion.}
\label{fig:FlowDiffRawData}
\end{figure}

Figure \ref{fig:FlowDiffRawData} shows plots of the power spectral density for an optical path-length near the edge of the channel (Fig. \ref{fig:FlowDiffRawData} (a)) and for an optical path-length corresponding to the center of the channel (Fig. \ref{fig:FlowDiffRawData} (b)). The circles show the data and the blue line shows the fitted model. The red line shows only the contribution of the flow term and the green line shows only the contribution of the diffusion term. It can be seen that at optical path-lengths with relatively low flow velocity values the decay of the power spectral density is well described by a Lorentzian (diffusion) decay. However, for relatively larger flow velocities, the decay of the power spectral density at low frequencies is well described by a Gaussian (flow) decay and at high frequencies it converges towards a Lorentzian decay.

\begin{figure}[t]
	\centering
		\includegraphics[width=0.7\textwidth]{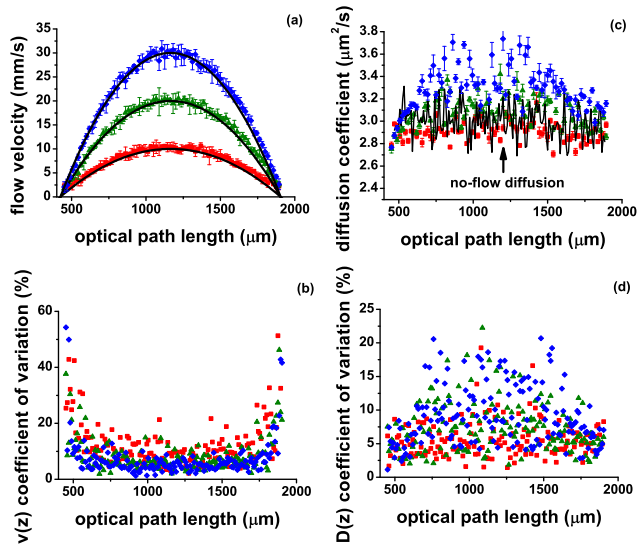}
\caption{(Color online) Simultaneous fitting of flow and diffusion for a set of different reference velocities: (a) Path-length resolved flow velocity. The solid line shows the reference velocity. (b) Coefficient of variation of the fitted flow velocity. (c) Path-length resolved diffusion coefficient for the velocities shown in (a). (d) Coefficient of variation of the fitted diffusion coefficient. For the sake of visualization, in Figs. (a) and (c) only every other data point has been plotted.}
\label{fig:FlowDiff}
\end{figure}

Figure \ref{fig:FlowDiff} shows results for the simultaneous measurement of the diffusion coefficient and the flow velocity based on data similar to the one shown in the previous figure. Figure \ref{fig:FlowDiff}(a) shows the path-length resolved flow velocities for a set of three varying reference velocities. The markers show the measured velocities and the lines show the reference velocities. As can be seen, the measured velocity values are well in agreement with the expected reference values. Figure \ref{fig:FlowDiff}(b) shows the corresponding coefficient of variation for the fitted velocities. For optical path-lengths close to the walls of the channel an increase in the coefficient of variation is observed similar to the case for a fit with fixed diffusion (cf. Fig. \ref{fig:TransFlow}). Figure \ref{fig:FlowDiff}(c) shows the path-length resolved diffusion coefficient corresponding to the flow velocities shown in Fig. \ref{fig:FlowDiff}(a). The markers show the measured diffusion coefficient and the line shows the diffusion coefficient measured in the no-flow case which have has used here as a reference. As can be observed, the diffusion coefficient measured at relatively low flow velocities is in good agreement with the reference values. For the relatively large flow velocities at the center of the flow channel, a path-length dependency of the measured diffusion coefficient is observed where the diffusion coefficient at the center of the channel is larger than the diffusion coefficient measured near the wall of the channel. As the maximum flow velocity through the center of the channel increases, the overestimation of the diffusion coefficient increases as well. Figure \ref{fig:FlowDiff}(d) shows the coefficient of variation for the fitted diffusion coefficient. For optical path-lengths close to the center of the channel an increase in the coefficient of variation is observed. 

\begin{figure}[t]
	\centering
		\includegraphics[width=0.7\textwidth]{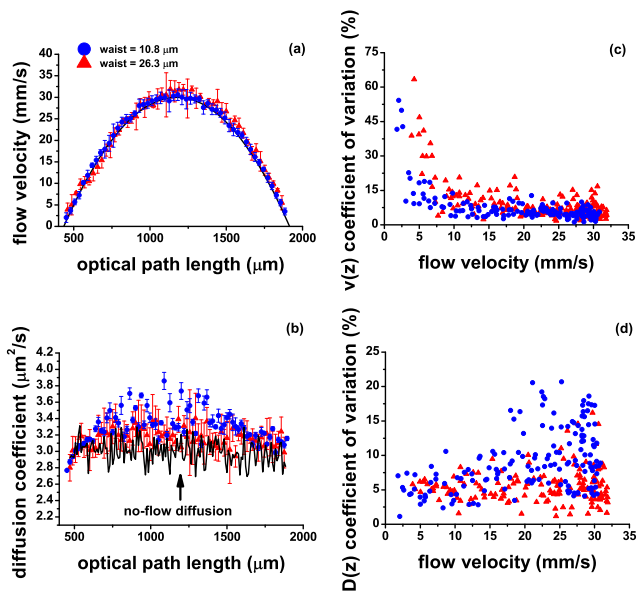}
\caption{(Color online) Simultaneous measurement of flow and diffusion at high flow rates. (a) Path-length resolved flow velocity measured using a waist of 10.8 \textmu m (red circles) and measured using a waist of 26.3 \textmu m (blue triangles). The solid line shows the reference velocity. (b) Path-length resolved diffusion coefficient corresponding to the velocity values shown in (a). The solid line shows the diffusion coefficient measured in the absence of flow. (c) Coefficient of variation for the flow velocities. (d) Coefficient of variation for the diffusion coefficient. For the sake of visualization, in Figs. (a) and (b) only every other data point has been plotted. Note that the data for the waist of $10.8$ \textmu m is the same as in Fig. \ref{fig:FlowDiff}.}
\label{fig:FlowDiff40vs100}
\end{figure}

Figure \ref{fig:FlowDiff40vs100} shows results for the simultaneous measurement of the diffusion coefficient and the flow velocity at high flow rates measured with two lenses resulting in a waist of $26.3$ \textmu m and a waist of $10.8$ \textmu m. Figure \ref{fig:FlowDiff40vs100}(a) shows the path-length resolved flow velocity and Fig. \ref{fig:FlowDiff40vs100}(b) shows the corresponding diffusion coefficient. As can be seen, the flow velocity measured with both lenses and the diffusion coefficient measured with the largest waist are in good agreement with the reference values. However, the diffusion coefficient measured with the smallest waist is overestimated in the center of the channel (cf. Fig. \ref{fig:FlowDiff}). Figure \ref{fig:FlowDiff40vs100}(c) and (d) show the coefficient of variation for the measured flow velocity and the diffusion coefficient, respectively. As can be seen, the coefficient of variation for the flow velocity measured with the smallest waist results in a reduced coefficient of variation for the low velocities when compared to the measurement with the largest waist. The coefficient of variation for the diffusion coefficient measured with the largest waist shows a reduced value for the high velocity range when compared to the measurements with the smallest waist.

\section{Discussion}
At hand of the theoretical model presented in Ref. \cite{Weiss2013} we have studied the simultaneous measurement of the local diffusion coefficient and flow velocity of a colloidal suspension. 

We have constrained the experimental and fitting conditions to study the influence of the diffusion coefficient, the flow velocity, and the beam waist on the measured OCT autocorrelation function and the estimated dynamical parameters. Our experimental results show that, although the diffusive and translation dynamics enter Eq. (\ref{eq:g}) independently, 
a reliable estimation of both quantities is challenging. 

Interestingly, interpreting the Gaussian (flow) term as being a transit time effect of the moving scatterers through the illumination profile would result in a dependence on the local beam radius. However, we have shown that the decay of the normalized OCT autocorrelation function is characterized by the beam waist irrespective of the position of the sample with respect of the focus. This is in excellent agreement with the theoretical prediction given in Ref. \cite{Taylor1986}, where this effect is attributed to the phase curvature of a focused Gaussian beam. More specifically, away from the focal plane the effect of beam radius increase and curvature increase compensate each other. As a result, the decay due to flow does not depend on the local Gaussian beam radius.

Our experimental results show that the uncertainty with which either the diffusion coefficient and the flow velocity can be quantified is determined by a competition between the single exponential (diffusion) and Gaussian (flow) decay terms in the autocorrelation function. In the particular case of relatively low flow velocities the decay of the autocorrelation function is dominated by the diffusive dynamics and is well described by a single exponential decay. In this case, the contribution of the Gaussian (flow) term to the total decay is small and therefore the estimation of the flow velocity suffers of a large uncertainty. In the case of relatively high flow velocities the decay of the autocorrelation function is dominated by the translational dynamics and is well described by a Gaussian decay. This results in a large uncertainty of the diffusion coefficient since the contribution of the single exponential decay to the total decay is relatively small. 

\begin{figure}[t]
	\centering
		\includegraphics[width=0.5\textwidth]{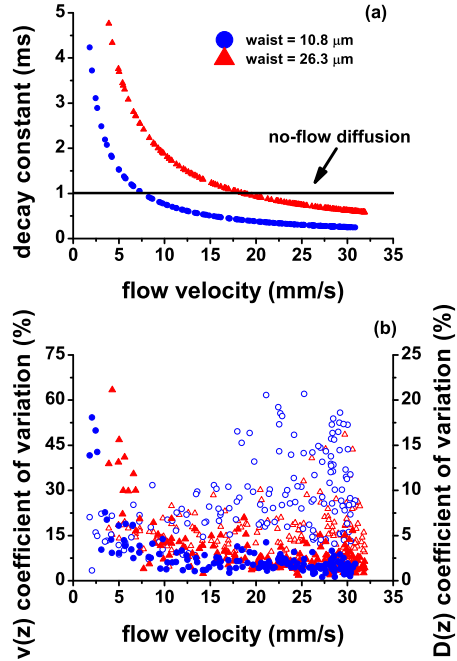}
\caption{(Color online) (a) Decay time constants of the flow (markers) and diffusion (line) terms of the OCT autocorrelation function. (b) Coefficient of variation for the flow velocity (filled markers, left ordinate) and for the diffusion coefficient (empty markers, right ordinate). The shape of the markers corresponds to the lenses shown in the legend in (a).}
\label{fig:diss1}
\end{figure}
The choice of the beam waist determines in which regime either dynamical process dominates the decay of the autocorrelation function. We identify two regimes for the choice of the beam waist. For applications where the focus lies in the estimation of the diffusion coefficient, it is more suitable to choose for a larger beam waist such that the decay due to diffusion dominates over a large part of the velocity range of interest. However, for applications where the focus lies in quantifying small flow velocities, a smaller beam waist is more suitable such that the decay due to flow dominates. As an illustrative example, we show in Fig. \ref{fig:diss1} the decay time constants $\tau_D = (2Dq^2)^{-1}$ (diffusion) and $\tau_v = (\sqrt{2}v/w)^{-1}$ (flow) and the corresponding uncertainties based on the data shown in Fig. \ref{fig:FlowDiff40vs100}. We can clearly see that as long as the decay due to diffusion dominates over the decay due to flow, the uncertainty of the diffusion coefficient remains close to $5 \%$. However, as the decay due to flow starts dominating, the uncertainty increases. A similar behavior is observed for the uncertainty of the flow velocity, where an uncertainty of $5\%$ is measured when the decay due to flow determines the total decay. Finally, by choosing either a large or a small beam waist we can shift the regime where either, the decay due to flow or due to diffusion, determines the total decay of the autocorrelation function.             

In the experiments performed for this work, we have constrained the flow to be perpendicular to the propagation direction of the imaging beam. By doing this, we have been able to study the influence of flow and diffusion on the amplitude terms of the normalized autocorrelation function. In essence, a generalization to arbitrary flow directions is possible. As we have shown previously, a longitudinal flow velocity component can be accurately measured by fitting an additional phase term to the normalized autocorrelation function \cite{Weiss2013}. For moderate shear rates the phase and the amplitude terms in the autocorrelation function remain uncoupled and thus, the results presented in this work can be generalized directly to the more complex case of arbitrary flow direction.   

In principle, a generalization of our results to arbitrary values of the beam waist is possible but should be performed with care. For small beam waists a strong reduction in the depth of focus would become a limitation for practical applications requiring long imaging ranges. Also, as a consequence of tight focusing a stronger dependency on the scattering angle is expected which has been shown to result in a more complex functional decay for the diffusive dynamics \cite{Roos1991}. Furthermore, a smaller beam waist will limit the maximum flow velocity that can be measured. In this case, the temporal resolution of the autocorrelation function should be sufficiently small to sample the flow decay $\tau_v = (\sqrt{2}v/w)^{-1}$ sufficiently. For large beam waists the influence of low signal-to-noise ratios \cite{Uribe2014} and multiple scattering \cite{Pine1988} should also be considered. A thorough description of these effects on the estimation uncertainty of the dynamical parameters is subject of future research.   

\section{Conclusion}
We presented measurements of the local diffusion coefficient and flow velocity based on the normalized autocorrelation function measured by optical coherence tomography. Based on our model, we have obtained accurate results by fitting the model to the measured data with only the physically relevant parameters as fitting parameters. Our results on a flowing solution of polystyrene spheres show that the flow velocity and the diffusion coefficient are reliably estimated. The regime where this is possible is determined by the diffusion coefficient of the sample, the local flow velocity, and the Gaussian beam waist produced by the focusing optics. We have experimentally shown that a smaller beam waist results in an improved estimation of the flow velocity and that a larger beam waist results in an improved estimation of the diffusion coefficient. Finally, we showed that the decay of the autocorrelation due to flow depends only on the Gaussian beam waist irrespective of the sample position with respect to the focus position.   
\section*{Acknowledgments}
We thank F.A.W. Coumans and P.R. Bloemen for discussions, and A.E. Grootemaat for characterizing the polystyrene particles. This work was financially supported by the IOP Photonic Devices program.


\begin{thebibliography}{99}
\bibitem{Orihara2011} H. Orihara, Y. Takikawa, ``Brownian motion in shear flow: Direct observation of anomalous diffusion,'' Physical Review E {\bf 84}, 061120 (2011).
\bibitem{Weber1993} R. Weber, R. Rambau, G. Schweiger, K. Lucas, ``Analysis of a flowing aerosol by correlation spectroscopy: concentration, aperture, velocity and particle size effects,'' Journal of Aerosol Science {\bf 24}, 485-499 (1993).
\bibitem{DiCarlo2007} D. {Di Carlo}, D. Irimia, R.G. Tompkins, M. Toner, ``Continuous inertial focusing, ordering, and separation of particles in microchannels,'' Proceedings of the National Academy of Sciences of the United States of America {\bf 104}, 18892-18897 (2007).
\bibitem{Joo2010} C. Joo, C.L. Evans, T. Stepinac, T. Hasan, J.F. {de Boer}, ``Diffusive and directional intracellular dynamics measured by field-based dynamic light scattering,'' Optics Express {\bf 18}, 2858-2871 (2010).
\bibitem{Drescher2013} K. Drescher, Y. Shen, B.L. Bassler, H.A. Stone, ``Biofilm streamers cause catastrophic disruption of flow with consequences for environmental and medical systems,'' Proceedings of the National Academy of Sciences {\bf 110}, 4345-4350 (2013).
\bibitem{Lee2013} J. Lee, H. Radharishnan, W. Wu, A. Daneshmand, M. Climov, C. Ayata, D.A. Boas, ``Quantitative imaging of cerebral blood flow velocity and intracellular motility using dynamic light scattering--optical coherence tomography,'' Journal of Cerebral Blood Flow \& Metabolism, {\bf 33}, 819-825 (2013).
\bibitem{Huang1991} D. Huang, E.A. Swanson, C.P. Lin, J.S. Schuman, W. G. Stinson, W. Chang, M.R. Hee, T. Flotte, K. Gregory, C.A. Puliafito, J.G. Fujimoto, ``Optical coherence tomography,'' Science {\bf 254}, 1178-1181 (1991).
\bibitem{Bizheva1998} K.K. Bizheva, A.M. Siegel, D.A. Boas, ``Path-length-resolved dynamic light scattering in highly random media: The transition to diffusing wave spectroscopy,'' Physical Review E {\bf 58}, 7664-7667 (1998).
\bibitem{Kalkman2010} J. Kalkman, R. Sprik, T.G. van Leeuwen, ``Path-length-resolved diffusive particle dynamics in spectral-domain optical coherence tomography,'' Physical Review Letters {\bf 105}, 198302 (2010).
\bibitem{Ishii2010} K. Ishii, T. Iwai, H. Xia, ``Hydrodynamic measurement of Brownian particles at a liquid-solid interface by low-coherence dynamic light scattering,'' Optics Express {\bf 18}, 7390-7396 (2010).
\bibitem{Chhetri2011} R.K. Chhetri , K.A. Kozek, A.C. Johnston-Peck, J.B. Tracy, A.L. Oldenburg, ``Imaging three-dimensional rotational diffusion of plasmon resonant gold nanorods using polarization-sensitive optical coherence tomography,'' Physical Review E {\bf 83}, 040903 (2011).
\bibitem{Huang2014} B.K. Huang, M.A. Choma, ``Resolving directional ambiguity in dynamic light scattering-based transverse motion velocimetry in optical coherence tomography,'' Optics Letters {\bf 39}, 521-524 (2014).
\bibitem{Uribe2014} N. Uribe-Patarroyo, M. Villiger, B.E. Bouma, ``Quantitative technique for robust and noise-tolerant speed measurements based on speckle decorrelation in optical coherence tomography,'' Optics Express {\bf 22}, 24411-24429 (2014).
\bibitem{Lee2012} J. Lee, W. Wu, J.Y. Jiang, B. Zhu, D.A. Boas, ``Dynamic lights scattering optical coherence tomography,'' Optics Express {\bf 20}, 22262-22277 (2012).
\bibitem{Weiss2013} N. Weiss, T.G. van Leeuwen, J. Kalkman, ``Localized measurement of longitudinal and transverse flow velocities in colloidal suspensions using optical coherence tomography,'' Physical Review E {\bf 88}, 042312 (2013).
\bibitem{Hendargo2011} H.C. Hendargo, R.P. McNabb, A.H. Dhalla, N. Shepherd, J. A. Izatt, ``Doppler velocity limitations spectrometer-based versus swept source optical coherence tomography,'' Biomedical Optics Express {\bf 2}, 2175-2188 (2011).
\bibitem{Arnaud1971} A. Arnaud, W.M. Hubbard, G.D. Mandeville, B. {de la Clavi\`ere}, E.A. Franke, J.M. Franke, ``Technique for fast measurement of Gaussian laser beam parameters,'' Applied Optics {\bf 10},  2775-2776 (1971).
\bibitem{Taylor1986} T.W. Taylor and C.M. Sorensen, ``Gaussian beam effects on the photon correlation spectrum from a flowing Brownian motion system,'' Applied Optics {\bf 25}, 2421-2426 (1986).
\bibitem{Roos1991} D.A. Ross, ``Focused laser beam effects in optical particle sizing by dynamic light scattering,'' Applied Optics {\bf 30}, 4882-4888 (1991).
\bibitem{Pine1988} A.J. Pine, D.A. Weitz, P.M. Chaikin, E. Herbolzheimer, ``Diffusing-Wave spectroscopy,'' Physical Review Letters {\bf 60}, 1134-1137 (1988).
\end{thebibliography}
\end{document}